\documentclass[a4paper,11pt]{article}
\usepackage{jcappub}
\usepackage{subcaption}
\usepackage{orcidlink}

\title{\boldmath Geometric Constraints on Quantum Gravity--Inspired Dispersion Relations}

\author{Ginés R. Pérez Teruel\,\orcidlink{0000-0001-8363-1366}}
\affiliation{Consellería de Educación, Cultura, Universidades y Empleo, Ministerio de Educación y Formación Profesional, Spain}

\emailAdd{gines.landau@gmail.com}

\abstract{
Modified dispersion relations (MDRs) arise in many quantum-gravity approaches, 
often in non-polynomial or non-analytic form beyond the reach of effective 
field theory (EFT). Logarithmic, exponential and trigonometric MDRs appear in 
causal set theory, nonlocal gravity and $\kappa$-Poincar\'e models, while Loop 
Quantum Gravity (LQG) yields polymeric (sine), holonomy, inverse-triad and 
semiclassical corrections.  
Using the geometric framework of Ref.~\cite{GRP}, we analyse the intrinsic 
curvature of the associated energy--momentum surfaces, where negative 
curvature ensures hyperbolic and stable propagation, and curvature sign changes 
or critical points indicate kinematical instabilities or new invariant scales.  

We apply this method exhaustively to all major MDRs derived in LQG and find 
that they remain strictly hyperbolic in the entire phenomenologically relevant 
regime, with no elliptic patches or critical branching.  
The same framework provides universal constraints on representative 
logarithmic, exponential and trigonometric MDRs beyond EFT.  
Thus, geometric criteria yield a unified and coordinate-independent assessment 
of stability, thresholds and invariant scales, and demonstrate the robustness 
of MDRs emerging from LQG.}

\begin{document}
\maketitle
\flushbottom

\section{Introduction}
\label{sec:intro}

Quantum gravity phenomenology often parameterizes departures from Special 
Relativity through modified dispersion relations (MDRs). Most analyses focus on 
polynomial corrections of the form
\begin{equation}
E_a^{2}=p^{2}+m_a^{2}+\sum_{n}\kappa_{a,n}\frac{p^{n}}{M^{\,n-2}},
\end{equation}
where $a$ labels the particle species and $\kappa_{a,n}$ are
dimensionless coefficients encoding possible LIV (and, in general,
species-dependent) corrections. These MDRs arise naturally in effective field theory (EFT) expansions and have been 
tightly constrained by astrophysical and cosmological observations 
\cite{Jacobson:2005bg,AmelinoCamelia:2013gna,Hossenfelder:2012jw,
Liberati:2013xla,Collins:2004bp}. 
EFT provides a systematic and model-independent parametrization of 
higher-dimension operators, making it extremely powerful for polynomial MDRs.

However, several approaches to quantum gravity predict MDRs that lie outside 
the polynomial EFT regime. Examples include logarithmic deformations in causal 
set theory and asymptotic safety \cite{Sorkin:2009,Henson:2009,Reuter:2012id}; 
exponential form factors in nonlocal and infinite-derivative gravity 
\cite{Biswas:2011ar,Modesto:2011kw}; trigonometric deformations in polymer 
quantization and $\kappa$-Poincar\'e models 
\cite{Ashtekar:2002sn,Hossain:2010eb,AmelinoCamelia:2002wr,
KowalskiGlikman:2004qa}; and both polynomial and non-polynomial MDRs suggested 
by Loop Quantum Gravity (LQG) 
\cite{AmelinoCamelia:2004xx,Gambini:1998it}.  
In addition, Doubly Special Relativity (DSR) introduces factorizable MDRs that 
preserve the geometry of special relativity but encode Lorentz symmetry through 
nonlinear transformations.

Standard EFT techniques, while extremely successful for polynomial MDRs, are 
not designed to handle the non-polynomial, non-analytic, or factorizable 
structures that arise in these quantum-gravity scenarios. This generates a gap 
between theoretical predictions and phenomenological tests: although 
high-energy astrophysical data probe momentum scales inaccessible in the 
laboratory, there is no general framework to translate these observations into 
constraints on MDRs that fall outside the EFT-polynomial class.

In this work, we close this gap by employing a unified geometric formulation in 
which any MDR is represented as an embedded surface in energy--momentum space. 
Geometric invariants of this surface---such as its slope, Gaussian curvature, 
and critical points---provide coordinate-independent diagnostics of stability, 
hyperbolicity, and the emergence of new invariant momentum scales. 
Three conceptually distinct kinematical conditions,
(i) hyperbolic and well-posed propagation, 
(ii) absence of spurious critical points, and 
(iii) threshold conditions for processes such as $\gamma\to e^+e^-$,
thus arise as simple geometric properties of a single surface in $\mathbb{R}^3$.

This unifying picture allows us to treat polynomial, non-polynomial, and 
factorizable MDRs on the same footing, and to derive 
model-independent constraints beyond EFT.  
A central result of this paper is the exhaustive application of this geometric 
method to the full spectrum of MDRs arising in Loop Quantum Gravity, including 
polymeric (sine), holonomy-corrected, inverse-triad and semiclassical/DSR-like 
forms. As we show below, all these MDRs remain entirely within the hyperbolic sector
of the off-shell surface throughout the phenomenologically
relevant regime, demonstrating the 
robustness of LQG kinematics when viewed through intrinsic geometric criteria. It is worth emphasizing, however, that the present framework should be understood primarily as a geometric classification of modified dispersion relations. While the intrinsic curvature and critical structure of the associated off--shell surface provide coordinate--independent assessments of hyperbolicity and stability, the interpretation of these geometric properties as physical viability conditions still requires an additional physical and phenomenological input. In particular, for sufficiently general non-polynomial or non-analytic dispersion relations, the global identification of curvature sign changes or critical structures may itself become algorithmically nontrivial. The framework developed here therefore separates geometric definability from practical computability, while retaining a unified mathematical language applicable beyond the standard EFT regime.

\section{Geometric framework}
\label{sec:geomframework}

We build on the geometric formalism developed in Ref.~\cite{GRP},
but emphasize a crucial distinction that is essential in our approach:
the modified dispersion relation is represented \emph{off--shell} by an
embedded surface in $\mathbb{R}^3$, while the physical mass shell corresponds
to the intersection with a fixed plane.

\paragraph*{Off--shell embedding.}
Given a function
\begin{equation}
  f(E,p)=E^2 - p^2 - m^2 - g(E,p),
  \label{eq:f-general}
\end{equation}
we define the \emph{embedded off--shell dispersion surface}
\begin{equation}
  \mathcal S \;=\; \{ (E,p,z)\in\mathbb{R}^3 \;\mid\; z=f(E,p) \}.
\end{equation}
A convenient parametrization is
\begin{equation}
  \mathbf r(E,p) = (E,\; p,\; f(E,p)).
\end{equation}
No constraint is imposed on $f$: the surface $\mathcal S$ is the full
graph of the function $f(E,p)$ in $\mathbb{R}^3$.
The physical on--shell dispersion relation is obtained only at the end
of the analysis, as the level set
\begin{equation}
  \Gamma = \{(E,p)\in\mathbb{R}^2 \mid f(E,p)=0\}
        = \mathcal S \cap \{z=0\}.
\end{equation}
Thus the geometric quantities we compute (tangent vectors, curvatures)
refer to the \emph{off--shell} surface $\mathcal S$, of which the
mass shell $\Gamma$ is a curve.

\paragraph*{Intrinsic geometry of the off--shell surface.}
The tangent vectors of $\mathcal S$ are
\begin{equation}
  \mathbf r_E = (1,\,0,\,f_E), \qquad 
  \mathbf r_p = (0,\,1,\,f_p),
\end{equation}
with
\[
  f_E := \frac{\partial f}{\partial E}, \qquad
  f_p := \frac{\partial f}{\partial p}.
\]
The first and second fundamental forms of $\mathcal S$ follow directly,
and the Gaussian curvature is
\begin{equation}
  K(E,p)=
  \frac{f_{EE}\,f_{pp}-f_{Ep}^2}{\bigl(1+f_E^2+f_p^2\bigr)^2}.
  \label{eq:K-gauss}
\end{equation}

\paragraph*{Hyperbolicity and physical viability.}
The sign of $K$ encodes the local type of the embedded surface.
For the purposes of modified dispersion relations:
\begin{align}
  K < 0 
  &\quad\Longrightarrow\quad 
    \text{saddle geometry $\Rightarrow$ off--shell hyperbolicity}, 
    \label{eq:Kneg}\\[3pt]
  K > 0 
  &\quad\Longrightarrow\quad 
    \text{elliptic patch $\Rightarrow$ loss of hyperbolicity (instability)}, \\[3pt]
  f_E=f_p=0 
  &\quad\Longrightarrow\quad
    \text{critical point of $f$ (possible new invariant scale).}
\end{align}
Physically acceptable propagation requires that the mass shell $\Gamma$
lies entirely in regions of $\mathcal S$ with $K<0$.
This geometric criterion is not ad hoc: for linear wave equations,
hyperbolicity of the principal part of the PDE is equivalent to
existence of real characteristic cones, finite propagation speed, and
well--posedness of the Cauchy problem
\cite{CourantHilbert,Hormander}.
The principal symbol coincides with the dispersion relation in momentum
space; therefore, hyperbolicity of the PDE is geometrically equivalent
to the saddle character ($K<0$) of the embedded off--shell dispersion
surface near the mass shell. 

If $K(E,p)$ becomes positive, the surface develops an elliptic region:
characteristic directions cease to exist and the evolution problem is
ill--posed. The marginal case $K=0$ corresponds to flattening of the
surface, typically signaling a degeneracy of the symbol or the birth of
new local branches.\\
The conditions introduced above should therefore be regarded as formally well-defined features of the off--shell dispersion surface. Their role as viability criteria relies on the additional assumption that the relevant propagation properties of the underlying theory are faithfully encoded in the associated principal symbol. From this perspective, the present framework offers a coordinate--independent description of hyperbolicity and stability, although the final physical significance of these features may still depend on the specific dynamical realization of the MDR under consideration.

\paragraph*{Thresholds as tangency of mass shells.}
Beyond hyperbolicity, reaction thresholds appear naturally in this
language as \emph{tangency conditions} between different embedded
surfaces. Let 
\[
  f_i(E_i,p_i)=0,\qquad i=A,B,C,
\]
denote the mass shells of species $A\to B+C$.
At threshold, kinematics requires collinearity of momenta,
\[
  p_A=p_B+p_C,\qquad E_A=E_B+E_C,
\]
together with a common slope in the $(E,p)$-plane,
\begin{equation}
  \frac{dE}{dp}\Big|_{A}
  =\frac{dE}{dp}\Big|_{B}
  =\frac{dE}{dp}\Big|_{C}.
\end{equation}
In terms of the off--shell functions $f_i$, this is equivalent to
\begin{equation}
  \nabla f_A(E_A,p_A)\;\parallel\;
  \nabla f_B(E_B,p_B)\;\parallel\;
  \nabla f_C(E_C,p_C),
\end{equation}
i.e.\ the normals to the three embedded surfaces become parallel at the
contact point. Geometrically, a threshold is a point where the three
surfaces admit a common tangent direction before intersecting on--shell.
This viewpoint generalizes the usual algebraic threshold analysis and
remains valid for non-polynomial or non-analytic dispersion relations.

\section{Examples of MDRs}
\subsection{Logarithmic (causal sets / asymptotic safety)}
Logarithmic deformations arise in two distinct approaches: 
(i) causal set theory, where discreteness modifies the d’Alembertian, and 
(ii) asymptotic safety, where running couplings induce logarithmic terms in the graviton propagator~\cite{Sorkin:2009,Henson:2009,Reuter:2012id}. 
A representative parameterization is
\begin{equation}
f(E,p)=E^{2}-p^{2}-m^{2}-\beta\,p^{2}\log\!\Big(1+\frac{p^{2}}{\Lambda^{2}}\Big).
\end{equation}
The relevant derivatives are
\begin{equation}
f_p=-2p-\beta\!\left[2p\log\!\Big(1+\tfrac{p^{2}}{\Lambda^{2}}\Big)
+\frac{2p^{3}}{p^{2}+\Lambda^{2}}\right],
\end{equation}
and
\begin{equation}
f_{pp}=-2-\beta\!\left[
2\log\!\Big(1+\tfrac{p^{2}}{\Lambda^{2}}\Big)
+\frac{4p^{2}}{p^{2}+\Lambda^{2}}
+\frac{2p^{2}(p^{2}+3\Lambda^{2})}{(p^{2}+\Lambda^{2})^{2}}
\right].
\end{equation}
Hyperbolicity ($K<0$) requires $f_{pp}<0$, i.e.
\begin{equation}
\beta\!\left[
2\log\!\Big(1+\tfrac{p^{2}}{\Lambda^{2}}\Big)
+\frac{4p^{2}}{p^{2}+\Lambda^{2}}
+\frac{2p^{2}(p^{2}+3\Lambda^{2})}{(p^{2}+\Lambda^{2})^{2}}
\right] > -2.
\end{equation}
For $\beta>0$ the bracket is positive and hyperbolicity holds automatically; 
For $\beta<0$, however, hyperbolicity imposes a bound on $|\beta|$ that depends on the maximum momentum probed, ensuring that $f_{pp}$ does not change sign within $[0,p_{\max}]$.

\paragraph*{Sharp tangency bound (logarithmic MDR, photons).}
For the logarithmic photon MDR
\begin{equation}
E^2=p^2+\beta\,p^{2}\log\!\Big(1+\frac{p^2}{\Lambda^2}\Big)\,,
\label{eq:logMDR-photon}
\end{equation}
the photon acquires an ``effective invariant'' $m_{\rm eff}^2(p)$. 
Photon decay $\gamma\!\to e^+e^-$ is kinematically allowed once the photon shell becomes tangent to the pair shell, i.e.
\begin{equation}
m_{\rm eff}^2(p_{\rm thr})=(2m_e)^2 
\quad\Longleftrightarrow\quad
\beta\,p_{\rm thr}^{2}\log\!\Big(1+\frac{p_{\rm thr}^2}{\Lambda^2}\Big)=4m_e^2.
\label{eq:log-threshold-eq}
\end{equation}
The non--observation of $\gamma\to e^+e^-$ up to the maximum detected momentum $p_{\max}^{(\gamma)}$ implies $p_{\rm thr}>p_{\max}^{(\gamma)}$, which yields the exact inequality
\begin{equation}
\beta \;<\;\frac{4m_e^2}{\big(p_{\max}^{(\gamma)}\big)^{2}\,
\log\!\big(1+{p_{\max}^{(\gamma)}}^{2}/\Lambda^2\big)}\,,\qquad (\beta>0).
\label{eq:beta-exact}
\end{equation}
In the regime $p_{\max}^2\ll \Lambda^2$ (valid for large $\Lambda$) the approximation $\log(1+x)\simeq x$ holds, and \eqref{eq:beta-exact} reduces to
\begin{equation}
\Lambda\;\gtrsim\;\frac{\sqrt{\beta}\,\big(p_{\max}^{(\gamma)}\big)^{2}}{2\,m_e}\,.
\label{eq:Lambda-approx}
\end{equation}

\noindent\emph{Numerical estimate.} 
For $p_{\max}^{(\gamma)}=100~{\rm TeV}=10^{5}~{\rm GeV}$, 
$m_e=0.511~{\rm MeV}=5.11\times10^{-4}~{\rm GeV}$, and $\beta\sim 1$,
\begin{equation}
\Lambda \;\gtrsim\; \frac{(10^{5})^{2}}{2\times 5.11\times10^{-4}}\ {\rm GeV}
\;\approx\;1\times 10^{13}\ {\rm GeV}.
\end{equation}
For photons of $200$--$300$~TeV the bound scales as $p_{\max}^{2}$ and naturally pushes $\Lambda$ into the $10^{13}$--$10^{14}$~GeV range.

\noindent\emph{Remarks.} 
(i) The strong phenomenological limit originates from the \emph{tangency criterion} (threshold), not from hyperbolicity $K<0$, which only constrains the $\beta<0$ branch. 
(ii) For $\beta>0$ photon decay is the dominant constraint; for $\beta<0$, curvature stability plays that role.

\subsection{Exponential (nonlocal / infinite-derivative gravity)}
In nonlocal and infinite-derivative gravity, the graviton propagator acquires entire-function form factors, typically exponential~\cite{Biswas:2011ar,Modesto:2011kw}. 
This translates into MDRs of the type
\begin{equation}
f(E,p)=E^{2}-p^{2}-m^{2}-\mu^{2}\Big(e^{p^{2}/M^{2}}-1\Big).
\end{equation}
One finds
\begin{equation}
f_{pp}=-2-\frac{2\mu^{2}}{M^{2}}\,e^{p^{2}/M^{2}}\!\left(1+\frac{2p^{2}}{M^{2}}\right).
\end{equation}
Hyperbolicity is always preserved ($f_{pp}<0$), but critical points can appear if $f_p=0$. 
The absence of such branching points below $p_{\max}$ constrains the parameters $(\mu,M)$.
\paragraph*{Photon decay bound.}
Although we already mentioned that exponential MDRs are often motivated for gravitons, here we apply the same functional form to photons as a phenomenological test case, since high-energy photon data provide the most stringent available probes of such deformations.
The photon dispersion relation ($m=0$) reads
\begin{equation}
E^{2}=p^{2}+\mu^{2}\Big(e^{p^{2}/M^{2}}-1\Big).
\end{equation}
This can be interpreted as the photon acquiring a momentum--dependent 
effective mass squared
\begin{equation}
m_{\rm eff}^{2}(p)=\mu^{2}\Big(e^{p^{2}/M^{2}}-1\Big).
\end{equation}

The decay $\gamma\to e^{+}e^{-}$ becomes kinematically possible once 
$m_{\rm eff}^{2}(p)\ge (2m_{e})^{2}$. 
Requiring that no such tangency occurs up to the highest observed photon momentum $p_{\rm obs}$ 
implies
\begin{equation}
\mu^{2}\Big(e^{p_{\rm obs}^{2}/M^{2}}-1\Big) < 4m_{e}^{2}.
\end{equation}
For $p_{\rm obs}\ll M$ this yields the approximate bound
\begin{equation}
M > \frac{\mu\,p_{\rm obs}}{2m_{e}}.
\end{equation}

Numerically, taking $p_{\rm obs}=100~{\rm TeV}=10^{5}\,{\rm GeV}$ 
and $m_{e}=5.11\times 10^{-4}\,{\rm GeV}$, one finds
\begin{equation}
M \gtrsim 10^{8}\,\mu \ {\rm GeV},
\end{equation}
which strengthens to $M\gtrsim 3\times 10^{8}\,\mu$ GeV for 
$p_{\rm obs}\sim 300$ TeV photons. 
Thus, exponential MDRs with $\mu\sim\mathcal O(1)$ are excluded unless 
the nonlocality scale $M$ lies well above the $10^{8}$--$10^{9}$ GeV range.

\subsection{Trigonometric (polymer quantization / $\kappa$-Poincaré)}

Polymer quantization, as well as $\kappa$-Poincaré deformations, lead to trigonometric or periodic MDRs~\cite{Ashtekar:2002sn,Hossain:2010eb,AmelinoCamelia:2002wr}. 
A simple realization is
\begin{equation}
f(E,p)=E^{2}-\frac{1}{\lambda^{2}}\sin^{2}(\lambda p)-m^{2}.
\end{equation}
For photons $(m=0)$ this reduces to
\begin{equation}
E(p)=\frac{1}{\lambda}|\sin(\lambda p)|,
\end{equation}
which is bounded above by $E(p)\leq 1/\lambda$.  
The very observation of photons with $E_{\rm obs}$ therefore requires the trivial reachability bound
\begin{equation}
\lambda \;\leq\; \frac{1}{E_{\rm obs}}.
\label{eq:bound_reach}
\end{equation}

In addition, the derivatives
\begin{equation}
f_p=-\tfrac{1}{\lambda}\sin(2\lambda p),\qquad f_{pp}=-2\cos(2\lambda p),
\end{equation}
provide further information. The condition $f_p=0$ gives extrema of $E(p)$ at
\begin{equation}
p_{\rm ext}=\frac{n\pi}{2\lambda},\qquad n\in\mathbb{Z}.
\end{equation}
Requiring that no extremum lies in the observational domain $[0,p_{\max}]$ enforces
\begin{equation}
\lambda < \frac{\pi}{2p_{\max}}.
\label{eq:bound_extrema}
\end{equation}
Meanwhile, the sign of $f_{pp}$ controls the Gaussian curvature: $f_{pp}<0$ corresponds to hyperbolic patches, $f_{pp}>0$ to elliptic ones.  
Avoiding elliptic domains up to $p_{\max}$ requires that the first zero of $\cos(2\lambda p)$, at $p=\pi/(4\lambda)$, lies above $p_{\max}$, leading to the stronger condition
\begin{equation}
\lambda < \frac{\pi}{4p_{\max}}.
\label{eq:bound_hyperbolicity}
\end{equation}

Numerically, for photons observed at $E_{\rm obs}\sim p_{\max}\sim 100\,{\rm TeV}$, the reachability bound \eqref{eq:bound_reach} gives $\lambda\lesssim 1.0\times10^{-5}\,{\rm GeV}^{-1}$, 
the no-extrema condition \eqref{eq:bound_extrema} yields $\lambda\lesssim 1.6\times10^{-5}\,{\rm GeV}^{-1}$, 
and the hyperbolicity bound \eqref{eq:bound_hyperbolicity} strengthens this to $\lambda \lesssim 7.9\times10^{-6}\,{\rm GeV}^{-1}$.

\begin{figure}[t]
    \centering
    \includegraphics[width=0.68\textwidth]{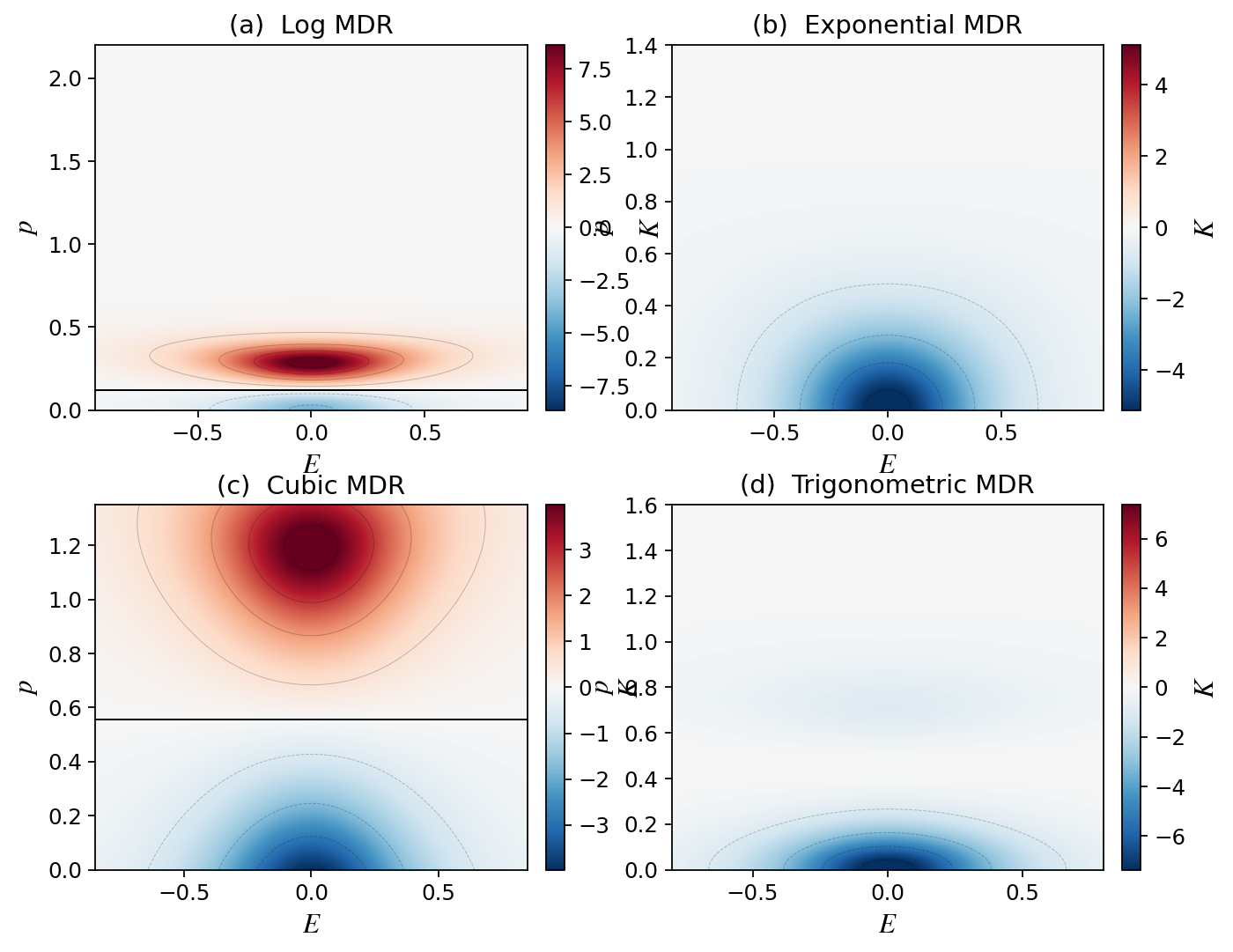}
    \caption{
    Gaussian curvature $K(E,p)$ of representative logarithmic, exponential, cubic, and trigonometric MDRs. Red regions indicate elliptic domains ($K>0$), associated with loss of hyperbolicity, while blue regions mark hyperbolic domains ($K<0$) corresponding to stable propagation.Each panel shows the onset of curvature sign changes and potential critical structures characteristic of each MDR class.}
    \label{fig:Kmaps}
\end{figure}

In this class of MDRs the improvement over the trivial reachability constraint is only by a factor of order unity, not by orders of magnitude. The added value of the geometric-shell approach here is therefore mainly conceptual: it unifies in a single framework the reachability condition, the absence of extrema, and the requirement of global hyperbolicity, providing a transparent rationale for why trigonometric MDRs cannot introduce new invariant scales below the observationally probed momenta.
The qualitative behaviour of the Gaussian curvature for these
representative non-polynomial MDRs is summarized in
Fig.~\ref{fig:Kmaps}. Logarithmic and trigonometric deformations
develop elliptic patches ($K>0$) separated from the hyperbolic domain
by $K=0$ curves, while exponential and cubic MDRs exhibit a purely
hyperbolic region over the observationally relevant range. These
curvature maps make explicit where new invariant scales or potential
instabilities may arise.

\section{Modified dispersion relations in Loop Quantum Gravity}
\label{sec:LQG-MDR}

LQG and its effective symmetry-reduced version,
Loop Quantum Cosmology (LQC), generically produce MDRs with structures that differ substantially from those of 
local effective field theories. 
These MDRs arise from the fundamental discreteness of quantum geometry 
via holonomies, polymerization of canonical variables, and inverse-triad 
operators. Crucially, many of these relations are \emph{non-polynomial} and 
\emph{non-analytic}, so EFT expansions in powers of $p/M_{\mathrm{Pl}}$ 
or $E/M_{\mathrm{Pl}}$ are not applicable.  
Our geometric framework---based on the off--shell embedded surface 
$\mathcal S = \{(E,p,z)\mid z=f(E,p)\}$---is therefore ideally suited 
to analyze the viability of these MDRs without relying on polynomial expansions.

Following Ref.\cite{GRP}, given a general dispersion function $f(E,p)$, we describe the \emph{off--shell} surface
\[
\mathcal S: \quad 
\mathbf r(E,p) = (E,\, p,\, f(E,p)),
\]
whose intrinsic geometry is encoded in the Gaussian curvature given by (\ref{eq:K-gauss})
\[
K(E,p) = \frac{f_{EE}f_{pp} - f_{Ep}^{2}}
              {\bigl(1+f_E^2+f_p^2\bigr)^2}.
\]
The physical dispersion law corresponds to the \emph{on--shell} curve
\[
\Gamma = \mathcal S \cap \{z=0\}
      = \{(E,p)\mid f(E,p)=0\}.
\]
Thus hyperbolicity, critical points, and thresholds are assessed through 
the off--shell geometry of $\mathcal S$, with the on--shell dynamics 
being restricted to the slice $z=0$.

The viability criteria are:
\begin{equation*}
\begin{aligned}
K<0 
  &\;\Longrightarrow\;
    \text{saddle geometry $\Rightarrow$ off--shell hyperbolicity},\\[4pt]
K>0 
  &\;\Longrightarrow\;
    \text{elliptic patch $\Rightarrow$ instability (loss of hyperbolicity)},\\[4pt]
f_E=f_p=0 
  &\;\Longrightarrow\;
    \text{critical point of $f$ (possible new invariant scale).}
\end{aligned}
\end{equation*}

We now apply this geometric analysis to the four principal families of
LQG-motivated MDR.

\subsection{Polymeric MDR with trigonometric holonomies}

Polymeric quantizations, following the holonomy-flux algebra of LQG,
replace the canonical momentum by the bounded operator
$p \mapsto \lambda^{-1}\sin(\lambda p)$.
This yields the off--shell dispersion function
\begin{equation}
f(E,p)
= E^{2} - \lambda^{-2}\sin^{2}(\lambda p) - m^{2},
\end{equation}
whose graph defines the surface 
$\mathcal S = \{(E,p,z)\mid z=f(E,p)\}$.
Such MDR appear in background-independent polymer quantizations of 
scalar fields and in LQC perturbation theory 
\cite{HossainHusainSeahra2009,GrainBarrau2010,BojowaldHossain2008,AgulloSinghReview}.

We have
\[
f_E = 2E,\quad f_{EE}=2,\qquad
f_p = -\tfrac{2}{\lambda}\sin(\lambda p)\cos(\lambda p),\quad
f_{pp}=-2\cos(2\lambda p),
\]
leading to the Gaussian curvature
\[
K(E,p)
= -\frac{4\cos(2\lambda p)}
       {\bigl(1+4E^{2}+f_p^{2}\bigr)^2}.
\]
In the experimentally accessible regime $|\lambda p|\ll 1$,
$\cos(2\lambda p)>0$ and therefore $K<0$: 
the off--shell surface is everywhere of saddle type near the on--shell curve.
The physical relation $\Gamma=\{f=0\}$ lies entirely within this hyperbolic region,
so no elliptic patches or instabilities arise.
Critical points would require $f_E=f_p=0$; for sub-Planckian $E,p$
no such point satisfies simultaneously the constraint $f(E,p)=0$.
Thus polymeric sine MDRs are fully viable in the hyperbolicity sense.

\subsection{Holonomy--induced MDR in the energy}

Holonomy corrections in the curvature or gravitational part of the 
Hamiltonian lead to higher-order, non-polynomial dependence on $E$.
A prototypical example is
\begin{equation}
f(E,p) = E^{2} - p^{2} - m^{2} - \alpha L_{\mathrm{Pl}}^{2}E^{4},
\end{equation}
arising in effective LQG and LQC models 
\cite{BrahmaEtAl2016,BojowaldPaily2012,GrainBarrau2010,BojowaldHossain2008}.
Its embedded surface $\mathcal S$ has
\[
f_E = 2E - 4\alpha L_{\mathrm{Pl}}^{2}E^{3},\qquad
f_{EE} = 2 - 12\alpha L_{\mathrm{Pl}}^{2}E^{2},
\]
\[
f_p=-2p,\qquad f_{pp}=-2.
\]
Hence
\[
K(E,p)
= -\,\frac{4\bigl(1-6\alpha L_{\mathrm{Pl}}^{2}E^{2}\bigr)}
        {\bigl(1+f_E^{2}+4p^{2}\bigr)^2}.
\]
For $E\ll M_{\mathrm{Pl}}$, the factor in parentheses is positive,
and the surface remains strictly hyperbolic.
The on--shell curve $f(E,p)=0$ does not intersect any $K>0$ region.
The only formal critical point ($f_E=f_p=0$) lies at $p=0$ and 
$E\sim 1/\sqrt{2\alpha}\,L_{\mathrm{Pl}}$, far above the physically relevant range
and incompatible with $f=0$ for realistic masses.
This type of MDR is therefore fully viable at sub-Planckian scales.

\subsection{Inverse-triad corrected MDR}

Quantization of inverse powers of the densitized triad introduces
non-analytic corrections in the kinetic term for fields or perturbations,
resulting in MDR of the type
\begin{equation}
f(E,p) = E^2 - p^{2}F(p) - m^{2},
\qquad
F(p)=1+\beta \frac{L_{\mathrm{Pl}}^{2}}{p^{2}}+\cdots,
\end{equation}
as found in several effective LQC analyses 
\cite{BojowaldLRR2008,AgulloSinghReview,WilsonEwing2017,
BojowaldHossain2008,GrainBarrau2010,BojowaldLQCObs2011,BojowaldCritique2020}.
For smooth positive $F$ in the relevant domain,
the second derivatives satisfy
\[
f_{EE}=2,\quad f_{pp} = -2F(p) - 4pF'(p) - p^{2}F''(p),
\]
In the phenomenologically accessible range, the sign of $f_{pp}$ remains negative along the portion of the embedded surface intersecting the mass shell. Consequently, the region traced by $\Gamma$ does not encounter elliptic patches within the momentum window of interest. Furthermore, no solutions of $f_E = f_p = 0$ coincide with $\Gamma$. Within this domain, inverse-triad MDR therefore show no indication of hyperbolicity loss or critical branching at sub-Planckian momenta.
\subsection{Semiclassical MDR and DSR-inspired expansions}

Semiclassical states of LQG and phenomenological deformations of the
Poincaré algebra often lead, at low orders, to MDR that can be written
as polynomial expansions of the form
\begin{equation}
f(E,p)=E^{2}-p^{2}-m^{2}\bigl(1+\gamma\,p/M_{\mathrm{Pl}}
                           +\delta\,p^{2}/M_{\mathrm{Pl}}^{2}
                           +\cdots\bigr),
\end{equation}
as discussed in 
\cite{AmelinoCameliaDSRPhenomenology,
Ling2006,Ling2010,
Girelli2012LQGPhenomenology,LiberatiDSR2007}.
These relations should be understood as \emph{EFT truncations} of more
fundamental structures (for instance, of exact DSR dispersion laws), and
are in general \emph{not} factorizable in the sense of
Eq.~\eqref{eq:DSR-factorizable}.

Within our framework they behave as generic, mildly deformed MDRs:
to the orders relevant for current observations one finds that the
off--shell curvature $K$ remains negative in the experimentally
accessible region, ensuring full hyperbolicity of $\mathcal S$
and consistency of the on--shell propagation. The case of exactly
factorizable DSR relations, which are geometrically equivalent to
special relativity, will be discussed separately in
Sec.~\ref{sec:DSR}.

Table~\ref{tab:LQG-MDR} summarizes the qualitative behavior 
of the main LQG-motivated MDR families.
In all cases, the on--shell curve $\Gamma$ lies fully within regions
of the off--shell surface $\mathcal S$ with $K<0$, guaranteeing 
hyperbolicity and absence of instabilities.

\begin{table}[t]
  \centering
  \begin{tabular}{lcccccc}
    \hline\hline
    MDR type & Schematic form & Analytic? & EFT? & Saturation & $K$ (IR) & Stable? \\
    \hline
    Polymeric (sine) 
      & $\lambda^{-2}\sin^{2}(\lambda p)$
      & No & No & Yes & $K<0$ & Yes \\
    Holonomy in $E$ 
      & $p^2+m^2+\alpha L_{\rm Pl}^2 E^4$
      & No & No & No & $K<0$ & Yes \\
    Inverse-triad 
      & $p^2F(p)$
      & No & No & No & $K<0$ & Yes \\
    Semiclassical / DSR 
      & $p^2+m^2(1+\gamma p/M_{\rm Pl}+\cdots)$
      & Yes & Yes & No & $K<0$ & Yes \\
    \hline\hline
  \end{tabular}
  \caption{
    Qualitative properties of representative LQG-motivated MDR.
    ``EFT'' refers to realizability in a finite-order local effective field
    theory. ``Saturation'' denotes bounded effective momentum as in the
    polymeric trigonometric case. The last row corresponds to semiclassical
    EFT truncations, not to the exact factorizable DSR relations
    discussed in Sec.~\ref{sec:DSR}.
  }
  \label{tab:LQG-MDR}
\end{table}

\begin{figure}[t]
  \centering
  \includegraphics[width=0.68\textwidth]{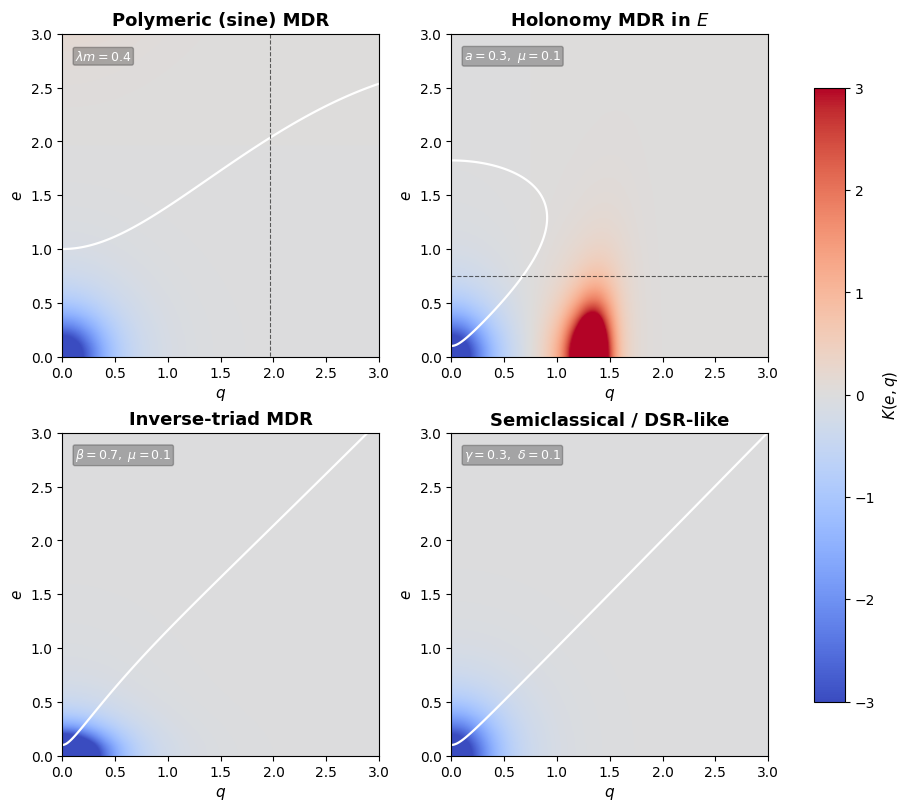}
  \caption{
  Gaussian curvature $K(E,p)$ for the four principal LQG–motivated modified
  dispersion relations: (a) polymeric (sine) MDR; (b) holonomy–induced MDR in the
  energy; (c) inverse–triad MDR; and (d) semiclassical/DSR-like MDR.
  The solid white curve indicates the physical mass shell $f(E,p)=0$,
  while the dashed black curve marks the locus $K=0$ separating hyperbolic
  ($K<0$) from elliptic ($K>0$) regions.
  In all cases the entire phenomenologically relevant part of the mass shell lies
  strictly within the hyperbolic domain, demonstrating the kinematical robustness
  of LQG-inspired MDRs at sub-Planckian energies.
  }
  \label{fig:LQG_Kmaps}
\end{figure}
The corresponding curvature profiles for these four LQG-inspired MDRs
are displayed in Fig.~\ref{fig:LQG_Kmaps}. In each case the physical
mass shell $f(E,p)=0$ (solid white curve) lies entirely within the
region with $K<0$, and never intersects the $K=0$ locus (dashed line)
within the phenomenologically accessed domain. This confirms
geometrically that LQG-motivated MDRs are strictly hyperbolic and free
of critical branching at sub-Planckian energies.

\section{Doubly Special Relativity}
\label{sec:DSR}

Doubly Special Relativity (DSR) modifies the kinematics of Special Relativity by
postulating, in addition to the invariant speed of light, an invariant energy
(or momentum) scale $M_{\rm Pl}$. Many realizations of DSR lead to modified
dispersion relations of the factorizable form ~\cite{AmelinoCamelia:2002wr,KowalskiGlikman:2004qa}
\begin{equation}
E^{2} f^{2}\!\left(\frac{E}{M_{\rm Pl}}\right)
 - p^{2} g^{2}\!\left(\frac{E}{M_{\rm Pl}}\right)
 = m^{2},
\label{eq:DSR-factorizable}
\end{equation}
with $f(0)=g(0)=1$.  In this case the dispersion function defining the
off--shell surface is
\begin{equation}
f_{\rm DSR}(E,p)
= E^{2} f^{2}\!\left(\frac{E}{M_{\rm Pl}}\right)
 - p^{2} g^{2}\!\left(\frac{E}{M_{\rm Pl}}\right)
 - m^{2}.
\end{equation}

\subsection*{Reparametrization to Special Relativity}

A key observation is that~\eqref{eq:DSR-factorizable} becomes the standard
relativistic form after a smooth, invertible change of variables.  Define
\begin{equation}
u(E) = E\,f\!\left(\frac{E}{M_{\rm Pl}}\right), 
\qquad 
v(E,p) = p\,g\!\left(\frac{E}{M_{\rm Pl}}\right).
\label{eq:DSR-map}
\end{equation}
Under this map, the dispersion relation becomes
\begin{equation}
u^{2} - v^{2} = m^{2},
\label{eq:u2v2}
\end{equation}
which is exactly the mass--shell of Special Relativity in $(u,v)$-coordinates.
Thus the \emph{on--shell curve}
$\Gamma = \{(E,p)\mid f_{\rm DSR}(E,p)=0\}$ is the image under the map
$\Phi:(E,p)\mapsto (u,v)$ of the standard hyperbola.

More importantly for our framework, the \emph{off--shell} surface
\[
\mathcal S_{\rm DSR} = \{ (E,p,f_{\rm DSR}(E,p))\}
\]
is mapped diffeomorphically to the surface
\[
\widetilde{\mathcal S}
= \{ (u,v,\, u^{2}-v^{2}-m^{2}) \},
\]
which is the \emph{graph} of the function
\[
\widetilde f(u,v)=u^{2}-v^{2}-m^{2},
\]
i.e.\ precisely the standard relativistic dispersion surface in the sense of
our geometric framework.

\subsection*{Intrinsic curvature is invariant under reparametrizations}

Since $(E,p)\mapsto (u,v)$ is smooth and invertible, it is a diffeomorphism
between coordinate charts of the two surfaces. Intrinsic geometric quantities
---and in particular, the Gaussian curvature---are invariant under
diffeomorphisms. Hence
\begin{equation}
K_{\rm DSR}(E,p) 
= K_{\rm SR}\bigl(u(E),v(E,p)\bigr).
\end{equation}

We now compute $K_{\rm SR}$ explicitly from the off--shell function
\[
\widetilde f(u,v) = u^{2}-v^{2}-m^{2}.
\]
From the general formula
\[
K=\frac{f_{uu}f_{vv}-f_{uv}^{2}}{(1+f_{u}^{2}+f_{v}^{2})^{2}},
\]
we obtain
\[
f_{u}=2u,\qquad f_{v}=-2v,
\qquad 
f_{uu}=2,\qquad f_{vv}=-2,\qquad f_{uv}=0.
\]
Therefore
\begin{equation}
K_{\rm SR}(u,v)
= \frac{(2)(-2)-0}{\bigl(1+4u^{2}+4v^{2}\bigr)^{2}}
= -\,\frac{4}{\bigl(1+4u^{2}+4v^{2}\bigr)^{2}}<0.
\label{eq:K-SR}
\end{equation}
The curvature is strictly negative everywhere: the SR surface is globally
hyperbolic and contains no elliptic patches or critical points.  Pulling it
back through the diffeomorphism~\eqref{eq:DSR-map}, we find
\begin{equation}
K_{\rm DSR}(E,p)
= -\frac{4}{\left(1+4u(E)^{2}+4v(E,p)^{2}\right)^{2}}
<0.
\end{equation}
Thus \emph{any factorizable DSR dispersion relation induces exactly the same
Gaussian curvature as Special Relativity, expressed in different coordinates.}

\subsection*{Consequences}

\begin{enumerate}
\item 
\emph{The geometry is the same as in SR.}
The off--shell surface $\mathcal S_{\rm DSR}$ is diffeomorphic to the SR
surface and inherits its everywhere-negative curvature. No new saddles, no
elliptic regions, and no critical points $(f_{E}=f_{p}=0)$ arise.

\item 
\emph{DSR kinematics is therefore a reparametrization of SR.}
What changes are the coordinate expressions of Lorentz transformations:
they act linearly on $(u,v)$ but become nonlinear in $(E,p)$ via the inverse
map $\Phi^{-1}$.

\item
\emph{DSR belongs to the ``trivial'' geometric class.}
Within our framework, factorizable MDRs (DSR, Gravity’s Rainbow) do not modify
the intrinsic geometry of the dispersion surface. In contrast, non-factorizable
MDRs (polynomial, logarithmic, exponential, trigonometric, polymeric) deform
the curvature, often producing new structures such as curvature sign changes
or genuine critical points.
\end{enumerate}

Hence DSR is recovered as a structurally trivial deformation of Special
Relativity in the geometric language: different coordinates, same intrinsic
geometry. This conclusion should nevertheless be understood in a strictly geometric sense. The fact that factorizable DSR relations are diffeomorphic to the special-relativistic dispersion surface implies that no new intrinsic curvature structures arise at the level of the off--shell geometry. However, nonlinear realizations of Lorentz symmetry may still modify the operational interpretation of energy--momentum variables, conservation laws, or multi-particle kinematics. The present analysis therefore isolates the intrinsic geometric content of DSR, without excluding additional physically nontrivial structures beyond the geometry of the dispersion surface itself.

\section{Phenomenological bounds}

The geometric framework developed above allows direct translation of
observational limits into model-independent constraints on non-polynomial
modified dispersion relations.  
High-energy astrophysical messengers probe momentum scales far beyond those
available in terrestrial experiments: photons up to $\sim 100$--$300$~TeV
(H.E.S.S., MAGIC, LHAASO), neutrinos up to the PeV range (IceCube), and
ultra--high--energy cosmic rays (UHECR) up to the EeV scale (Auger).  
Requiring that no instabilities, critical points, or decay thresholds occur
below these energies yields robust constraints on MDR parameters.

\begin{figure}[t]
  \centering
  \begin{subfigure}[t]{0.48\textwidth}
    \centering
    \includegraphics[width=\linewidth]{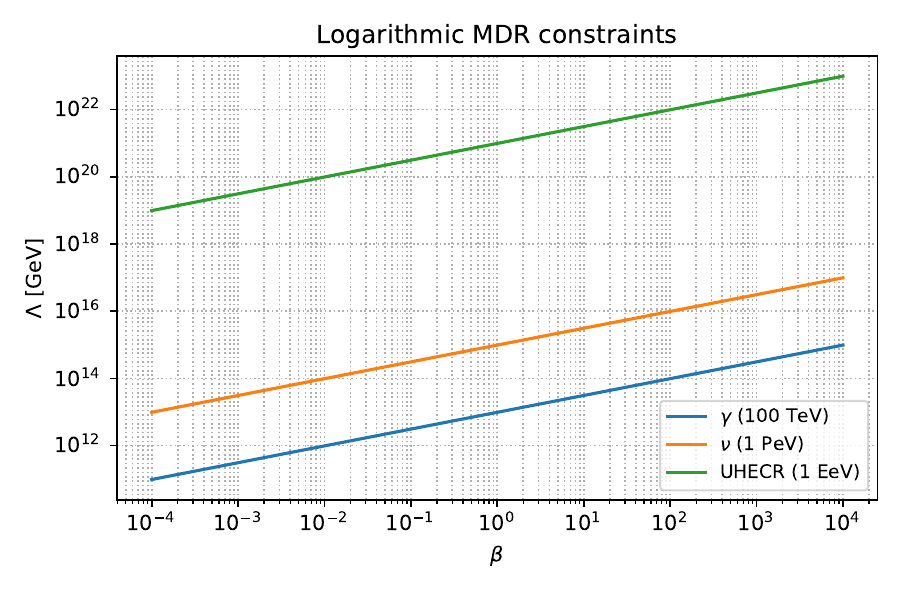}
    \caption{Logarithmic MDR}
    \label{fig:logMDR}
  \end{subfigure}
  \hfill
  \begin{subfigure}[t]{0.48\textwidth}
    \centering
    \includegraphics[width=\linewidth]{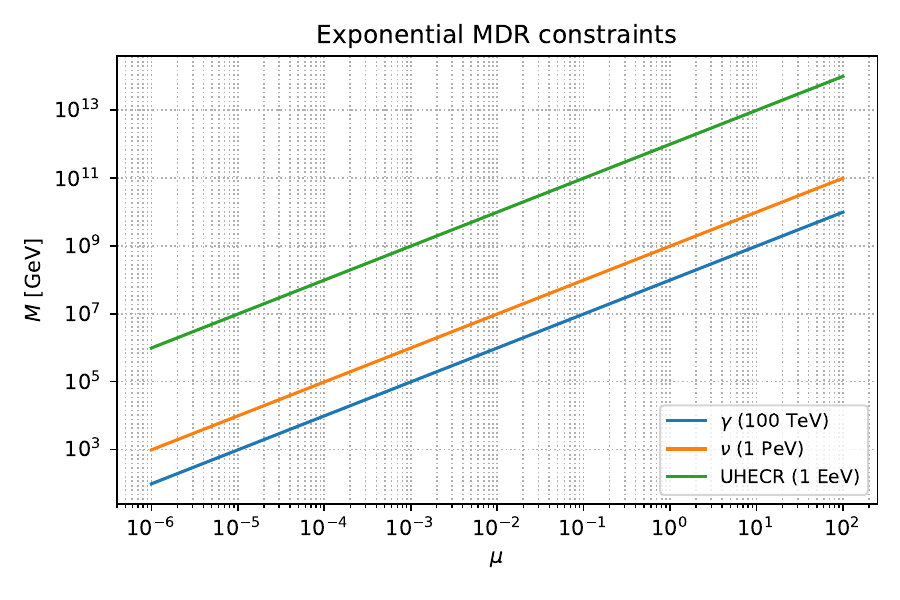}
    \caption{Exponential MDR}
    \label{fig:expMDR}
  \end{subfigure}

  \caption{Exclusion lines for logarithmic (a) and exponential (b) MDRs.
  For each messenger—photons ($E\!\lesssim\!100$--$300$~TeV), neutrinos
  ($E\!\lesssim\!1$~PeV), and UHECR ($E\!\lesssim\!1$~EeV)—the colored curve marks
  the threshold at which vacuum decay or related instabilities would already
  occur. Parameter values \emph{below} each curve are excluded; the region above
  the curves is phenomenologically viable.}
  \label{fig:log-exp}
\end{figure}

\paragraph*{Logarithmic MDRs.}
From the tangency condition (Eq.~12) the photon acquires an effective mass
$m_{\rm eff}^{2}(p)$ which must satisfy $m_{\rm eff}^{2}(p_{\rm obs}) < (2m_{e})^{2}$.
This results in a lower bound $\Lambda > \Lambda_{\min}(\beta)$.
For $\beta\simeq 1$ and $p_{\rm obs}=100$--$300$~TeV,
this gives $\Lambda \gtrsim 10^{13}$--$10^{14}$~GeV.
The allowed region lies above the exclusion lines in Fig.~\ref{fig:log-exp}.

\paragraph*{Exponential MDRs.}
For the nonlocal form factors considered in Sec.~3.2, the no-decay condition
(Eqs.~17--18) requires
$M > M_{\min}(\mu)$, with $M_{\min}$ determined by the photon, neutrino, or
UHECR thresholds.  
For $\mu\sim \mathcal O(1)$ and $p_{\rm obs}\sim 100$--$300$~TeV,
one finds $M \gtrsim 10^{8}$--$10^{9}$~GeV.  
Again, the phenomenologically allowed region corresponds to $M$ above the
curves in Fig.~\ref{fig:log-exp}.

\paragraph*{Trigonometric MDRs.}
Hyperbolicity, reachability, and absence of extrema (Eqs.~21--24) jointly require
$\lambda < \pi / (4 p_{\rm obs})$, implying
$\lambda \lesssim \text{few} \times 10^{-6}~{\rm GeV}^{-1}$  
for TeV--PeV photons.\\

These results---summarized in Fig.~\ref{fig:log-exp}--- are the
first model-independent bounds on non-analytic MDRs.  
They follow solely from geometric consistency (hyperbolicity, absence of
critical points, and threshold tangency), and therefore apply even when the
MDR cannot be approximated by any finite EFT expansion.

\section{Conclusions}

We have shown that modified dispersion relations can be analysed in a fully
general, coordinate-independent manner by studying the intrinsic geometry of
their associated off--shell energy--momentum surfaces.  
Within this framework, three seemingly distinct viability conditions—
(i) hyperbolicity and well-posed propagation,  
(ii) absence of critical points introducing new invariant scales, and  
(iii) threshold tangency for vacuum decays such as $\gamma\to e^{+}e^{-}$—  
all arise as simple geometric features of a single embedded surface.

Applying this method to non-polynomial MDRs motivated by causal sets,
nonlocal/infinite-derivative gravity, polymer quantization, and
$\kappa$-Poincaré models yields strong, model-independent bounds on
logarithmic, exponential, and trigonometric deformations. These constraints, shown in Fig.~\ref{fig:log-exp}, are
entirely geometric and therefore remain valid even when EFT expansions break
down.  
To our knowledge, these results constitute the first model-independent
bounds derived from intrinsic geometric consistency conditions for
non-analytic MDRs.

A central outcome of this work is the exhaustive geometric analysis of the
four main LQG-motivated MDRs.  
Polymeric (sine), holonomy-corrected, inverse-triad, and semiclassical/DSR-like
relations all remain strictly hyperbolic throughout the entire observational
window, with no elliptic patches or critical branching.  
Thus, within the geometric classification developed here, LQG predicts a class
of MDRs that are intrinsically stable and phenomenologically robust.\\

A further conceptual implication of the present framework is the distinction between geometric definability and algorithmic tractability. While the off--shell surface associated with a modified dispersion relation provides a precise geometric object from which hyperbolicity, threshold structure, and critical behavior may in principle be extracted, it is not guaranteed that these properties can always be decided by finite computational procedures for arbitrary non-polynomial dispersion functions. This suggests that, beyond the effective-field-theory regime, the classification of modified dispersion relations may encounter nontrivial computability limitations, even when the underlying geometric notions remain perfectly well defined. In this sense, the framework developed here may also be viewed as clarifying the boundary between geometrically meaningful and effectively computable kinematical structures.

In conclusion, geometric consistency of the off--shell surface provides a unifying criterion capable of constraining a wide variety of quantum-gravity
scenarios—polynomial and non-polynomial, analytic and non-analytic—well beyond the reach of standard EFT methods.  
The framework offers a bridge between quantum-gravity phenomenology
and high-energy observations, and naturally accommodates both genuinely new
deformations and reparametrization-invariant cases such as DSR.

\section*{Acknowledgements}
The author thanks the anonymous referees for their valuable comments and suggestions, which helped refine the conceptual framing and broaden the scope of the manuscript.

\bibliographystyle{apsrev4-2}

\end{document}